# On Being Prepared: Automated Vehicle Incident Management Exercise Practices

**Authors**
Laura Fraade-Blanar,* Jackson Graves,* Aaron Clark-Ginsberg, Sam Cooper, Jeremy Kaufman, Jackie Lindsey, Piet van Os
*Co-first authors

**Abstract**
Incident management (IM) has evolved over recent decades to cover an ever-expanding array of hazards and systems. Barring real-life experience, exercises are a key tool in developing an effective IM program. As part of enterprise resilience and operational readiness, IM practitioners design exercises to understand and build capabilities for efficiently and effectively responding to incidents. Combining research and policy on an emerging transportation technology, automated vehicles (AVs), with established practices for IM, we describe what makes exercises effective and how they can be used to identify gaps and develop capacity. A range of exercise types exist, from workshops, to tabletops, to drills, and to full-scale exercises. Every stage of an exercise — preparing, setting up, facilitating, closing, and assessing — is interconnected and should further the exercise's objectives. AV IM practitioners can maximize the utility of a mature exercise program through exercise quality, diversity, and volume. The need to respond to AV incidents, from collisions to natural hazards, is inevitable. This paper tailors existing practices to a series of frameworks, predicated on the unique socio-technical nature of AVs to provide first-of-its-kind guidance for AV IM exercises. Additionally, these frameworks are company-neutral to enable intercompany and inter-organization collaboration.

# 1. Introduction

The message can come in at any time. It could be a page at 10am on a Tuesday, or a phone call as Saturday night becomes Sunday morning. The content varies only slightly. “There has been an incident. Join this meeting immediately.” Select people quickly fill the room or video call, each representing specific responsibilities and expertises. The Incident Commander starts talking. It may be a collision with an automated vehicle, also known as a self-driving vehicle (AV)[1]. Or a cybersecurity attack, an earthquake, a facility fire, or any number of events. The bad news is the incident is severe. The good news is none of this is real: it is all an exercise, part of an AV company’s incident management (IM) program.

IM is the “the science of managing complex systems and multidisciplinary personnel to address emergencies or disasters, across all hazards, and through the phases of mitigation, preparedness, response, and recovery.” [1] A mature discipline, IM concepts and precepts are used in a range of industries [2,3,4]. In emerging technology, private companies’ IM programs can engender broad social benefits in safety, security, and wellbeing [2,5,6].

Exercises are a key component of an effective IM program. An exercise is “an event or activity, delivered through discussion or action, to develop, assess, or validate plans, policies, procedures, and capabilities that jurisdictions/organizations can use to achieve planned objectives.” [7] Exercises are used widely because they provide opportunity to rehearse and improve capabilities to respond to a variety of hazards [8]. Correctly done, exercises pressure test plans (processes and procedures), galvanize actions, identify weak spots, and build robust capabilities for responding to the expected and unexpected [9,10]. Ineffective exercises can result in limited benefits and may generate new risks [11,12]. In these situations, when a crisis strikes, organizations may be left less prepared rather than more so.

When applied to AVs, IM is the coordinated, structured process of responding to an event to immediately prioritize the safety of people, regulatory compliance, public trust, and business continuity [13,14]. AVs have the potential for involvement in the vast variety of incidents specific to transportation (e.g., crashes, unusual traffic events, etc.) in addition to those that may affect the transportation sector (e.g., natural hazards, civil unrest, etc.). The ability to respond to these incidents is thus a key part of operational readiness and recovery for AV organizations.

---

[1] AV refers to vehicles with an automated driving system, “the hardware and software that are collectively capable of performing the entire *DDT* (dynamic driving task) on a *sustained* basis.” [15] AVs span vehicle types, from heavy trucks to passenger vehicles to delivery bots and beyond. At this time, AVs generally operate through a fleet-based approach rather than individual private ownership (and as such, this paper primarily focuses on fleet-based models, although fundamental concepts may also apply to private ownership). Depending on an AV company’s business model and approach to validating its technology, a human affiliated with the fleet operator may or may not be physically present in the vehicle to assist with monitoring, provide feedback, or manually operate the vehicle.

This paper focuses on Level 4 and 5 AVs. For more information on the levels of automation, please refer to SAE J3016 [15].

Canonical IM sources provide limited guidance for AV operators. For example, the existing standard practice handbooks in the United States, such as FEMA's ICS NIMS (Incident Command System National Incident Management System) and HSEEP (Homeland Security Exercise and Evaluation Program), have much to offer AV IM practitioners. However, these documents were written for federal, state, local, and tribal government professionals in the United States. ISO guidance such as 22320 (Guidelines for Incident Management) is more audience-agnostic, but recommendations remain broad [13]. Translation is needed for these concepts to apply to a private enterprise operating in the AV space.

Neither is there considerable guidance from AV-centric Standards and Best Practices. UL 4600 section 10.6.1 focuses on the importance of reporting incidents and tracing contributing factors [16]. AVSC's Best Practices on First Responder Interactions (AVSC-I-01-2024) describes one area of IM activities, but does not relate these to a broader program [17]. AVSC's Best Practice on Post-Crash Response (AVSC-I-05-2025) describes incident response protocols as out of scope [18].

In technical literature, the focus is on AVs' potential use in improving disaster response in general or for a specific region or sub-group [19,20,21,22,23,24,25,26,27,28,29,30]. Suggestions span activities such as evacuation and supply delivery, surveillance, and government-requested creation of movable, vehicle-based blockades. These texts potentially provide important content to an ongoing conversation. But they do not provide guidance on the institutional systems and processes that make these activities possible.

Overall, as noted by an IM researcher, "planning and running exercises is in many ways a 'practitioner's game,' where organizations or small communities develop local practices without exchanging their knowledge with external parties, without benchmarking their approaches with others." [31] For AV IM, and for transportation IM broadly [32], there is limited information on exercise design and implementation [31].[2] This paper aims to address the scarcity of research and literature connecting the fields of incident management exercises and autonomous vehicles and provide a first-of-its-kind tailored, yet company-neutral guide for AV IM practitioners.[3] IM is a shared responsibility, requiring sustained coordination and collaboration across a diverse set of stakeholders [33,34,35]. By laying the groundwork for intercompany and inter-organization collaboration and providing actionable guidance currently lacking in the industry, this paper is a step toward turning a "practitioner's game" into a game that all AV companies and their partners can play together.

---

[2] A handbook for transportation professionals on exercises noted, "DHS has provided extensive general guidance on developing training and exercise programs for public entities (HSEEP Web n.d.), but little has been done to focus that material on the transportation sector specifically." Additionally, "'The consensus across all transit agencies interviewed was that there is a need to augment the HSEEP [Homeland Security Exercise and Evaluation Program] documents with practical guidance on exercise design, and exercise documentation development.'" [32]

[3] This article utilizes the expertises of the authors, which include backgrounds encompassing: emergency management, business continuity, maritime safety, firefighting, emergency medical services, IM, AV operations, public health, and over a quarter century in AV safety.

The importance of interoperability beyond company borders is one of the reasons that, of all the IM capabilities for which an AV practitioner must maintain diligence and expertise, we focus in this paper on exercises. As AV companies increase in maturity and scale within shared and separate operational areas, so does the importance of a standard vernacular and practice between them as well as partners and non-AV IM practitioners (e.g., local Departments of Emergency Management). The second reason is exercise importance; effective exercises serve as a method to cultivate and to assess readiness. Conversely, ineffective exercises instill false confidence and decouple preparedness plans from actual practice, resulting in a detrimental response when hazards strike[4] [12]. Third, exercises can be discussed in a way that is technology-, business-model-, location-, and vehicle type-agnostic, and applies at all stages of the safety determination lifecycle [36], including pre- and post-commercial deployment. Consequently, such work is salient across the industry.

We begin by describing the field of IM and exercises generally (section 2 and 2.1), then applying these concepts to AVs (section 3). Next we describe the role that exercises can play in developing an all-hazards approach to AV IM (section 3.1) and outline different exercise types (section 3.2). Continuing, stages of an exercise are detailed with activities linked to each of the established IM domains, ensuring coverage of all IM activities (section 4). We conclude with the limitations of exercises (section 5), and a discussion (section 6).

# 2. Incident Management

IM has been used for decades across incidents and industries. It originated in the 1970s [37], where firefighters needed to coordinate response across administrative boundaries and has since been applied to many other incident types such as: infectious disease outbreaks [38], oil spills [39], electric grid failures, cyber attacks [2], and AI loss of control incidents [40].

Many of these applications draw on the US's federal framework for incident management, ICS NIMS [41]. However, effective IM is not purely about alignment to established protocols. Policies are not substitutes for practice and may miss fundamental aspects of operations [42], overlap, or compete [43,44]. Taking a broader view, functional domain areas for effective IM (referred to as "domains" or "IM domains" in this paper) [45] include:

- situational awareness and information sharing;
- incident action and implementation planning;
- resource management;
- coordination and collaboration;
- feedback and continuous quality improvement.

[4] A vague scenario, for example, will limit the evaluator's ability to draw conclusions about the effectiveness and completeness of response procedures (assuming it is not the goal to assess how participants proceed under very limited and vague information).

## 2.1 Incident Management Exercises

An effective IM program involves establishing systems, protocols, and processes ahead of an incident that build capacity across functional IM domains. Engaging in these activities through exercises can improve IM [11,7].

There is scattered literature on what makes IM exercises effective. First exercises must closely simulate the diverse range of real-life scenarios, including using realistic settings and challenges relevant to the specific context of the participants [46,47,10]. Objectives must be clear, with exercises aligned with the desired dimension of preparedness and response capabilities [48,49]. Exercises are designed in line with the organization's broader resilience objectives, and can prepare personnel for the stresses, emotions, and cognitive load that inevitably accompany real-time response. By simulating near-real-world conditions, effective exercises provide participants with contained environments in which they can calibrate expectations and increase overall competence, confidence, and familiarity with the response process for a given scenario.

Planning is required, including templates, identified roles, responsibilities, and logistical considerations [49,50]. During the exercise, there is a need for active participants and teamwork [9,51] and, after the exercise concludes, after-action reports (AARs), focus groups, and other evaluations on strengths and weaknesses to learn and make programmatic adjustments [48,52,50]. Finally, work should focus on examining the exercise itself to support improving quality for future exercises [9]. How this occurs is inevitably context contingent, varying by the nature of the incident, the IM organization(s) involved, and the exercise aims and objectives. At the same time, standardization in exercises is necessary for ensuring interoperability in response and a common baseline of readiness.[5]

# 3. IM and exercises in the context of AV fleet operations

AV fleets represent a new form of technology which may be affected by hazards applicable to the broader transportation system, such collisions or intentional physical acts meant to interfere with an AV, as well as others not specific to transportation systems, including natural hazards such as earthquakes, floods, and wildfires.[6] And as a distributed infrastructure, AVs face these hazards across geographic regions and contexts. AVs and their constellation of networks, tools, and infrastructure also share risks assumed by other complex technological systems including cybersecurity and network outages. For this reason, an all-hazard approach[7] to AV IM, focused

---

[5] Interoperability is important due to the growing number of entities within the AV ecosystem including AV developers, fleet operators and governmental bodies. In an incident, no one works alone. Response hinges on collaboration between different specialties within an organization and, depending on the scale of the incident, across a range of different types of organizations.

[6] Discussion of the appropriate response within the context of AVs specific to each hazard is beyond the scope of this paper.

[7] This is "an approach to emergency management that addresses natural disasters and accidental or human-made events, including any natural catastrophe (e.g., hurricane, tornado, storm, high water, wind-driven water, tidal wave, tsunami, earthquake, volcanic eruption, landslide, mudslide, snowstorm, or

on building and supporting core competencies and institutional capacity for a range of incidents, may support resilience better than a hazard-specific approach [53,54].

An all-hazards-focused (discussed below in 3.1) AV IM program takes a holistic, activities-centric approach (Figure 1), focused on:

- **Prevention:** Working with cross-sectional internal and external teams to proactively identify and mitigate safety risks;
- **Preparation:** Assessing readiness of the organization for a wide range of risk types and planning, conducting, and evaluating training sessions and exercises that are designed to address identified risk types;
- **Response:** Responding to events as they unfold with the goal of minimizing impact to affected individuals and property;
- **Recovery:** Returning to normal operations as safely and as quickly as possible;
- **Improvement:** Conducting post-event and post-exercise reviews and implementing learnings from real and simulated events [53].

Figure 1: AV IM Program Activities

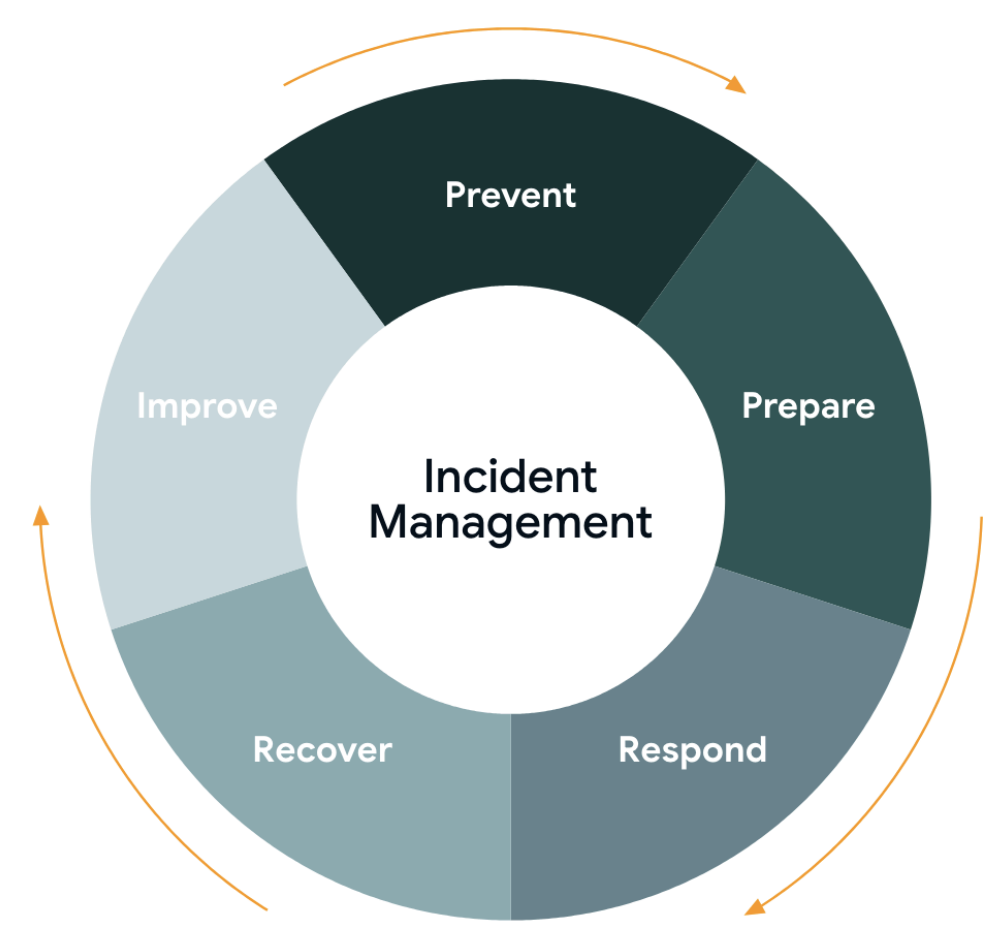

As this approach suggests, AV IM necessitates the seamless integration of proactive preparedness (including exercises) and reactive response functions. Each enables the success of the other.

Exercises allow IM teams to proactively practice reacting in a standardized and effective manner in order to minimize the impact of a given disruption. They are controlled environments for teams to make mistakes, test extreme scenarios, and refine complex, cross-functional workflows. This allows an organization to identify and address critical operational and capability defects and training gaps without incurring the high costs, public safety consequences, and reputational risk of failure during a real incident. Additionally, exercises can inform organizational resilience (i.e., what it takes for the final safeguard to fall). The structured learning process means that when a real incident strikes, the team executes a proven plan rather than being forced to engage in high-stakes, real-time trial-and-error.

The question of when an AV IM program begins exercises is one of maturity; simply put, an exercise program should begin when the entity has sufficient understanding of what to simulate, vis-a-vis incidents and applicable hazard types, and how to respond, vis-a-vis having at least the rudiments of response protocols for those incidents and hazard types. An AV company's

---

drought); fire; explosion; or other catastrophe, including those involving terrorist use of a weapon of mass destruction in any part of the United States that causes, or may cause, substantial damage or injury to civilian property or persons." [55]

exercise program may mature over time. A company's exercise program should continually iterate and account for unique factors that emerge as the organization matures and expands its operational design domain (ODD)[8], fleet size, and geographic footprint. A bespoke and experience-based program may mature to one more rooted in applicable portions of standardized exercise frameworks (e.g., HSEEP). This maturity coincides with an increased professionalization and specialization of AV IM practitioners and exercises.

An AV company's exercise cycle is likely to be most effective when administered by a centralized program staffed with dedicated, experienced practitioners that possess an in-depth understanding of the organization's technology and operations, range of applicable hazards, as well as the craft of preparing, conducting, and assessing exercises.

The cadence of exercises is influenced by a range of factors, but triggers can be roughly grouped into time-based and change-based. Within the former, a developer may, for example, choose to schedule regular full-scale test-track exercises to use the track efficiently and support advanced logistical planning. This could also include developing a time-based cycle of exercise types and scenarios as is common in other transportation industries such as aviation (e.g., O'Hare International Airport's triennial exercise program [56] in support of its Airport Emergency Plan [57]), bus, and rail (e.g., Metro Atlanta Regional Transit Authority [58] in support of Georgia's Program Standard Rail Transit Safety and Security [59]). For the latter (change-based triggers), an external incident, new regulation, new internal process, technological development, policy, tooling, or personnel change, or change in the transportation ecosystem, etc. may all warrant conducting a formative or educational exercise (as defined in section 3.2) to understand how this new change affects incident response protocols. As these examples suggest, triggers can arise from sources within, or outside of, the company.

Existing frameworks such as HSEEP set forth a strong foundation for exercise programs and progress in the stages of development. Still, exercises must factor in nuances specific to the AV industry as well as those of each AV company. One such nuance includes identifying the appropriate participant groups per exercise. As an AV company's operations increase in scale and complexity, they may benefit from conducting exercises with a variety of audiences ranging from internal cross-functional teams to external fleet management and operational partners and applicable government entities. The appropriate set of exercise participants will depend on a given AV company's business model, ODD, and exercise needs (as discussed in Section 4.1.1). While core response functions may remain centralized within an AV company, exercises with external partners and government agencies can help clarify expectations, roles, and responsibilities during an incident's initial detection, response, and resolution. Such exercises can be valuable tools in de-risking joint operations on a cadenced basis or in advance of operational milestones. For example, exercises with fleet management and operational partners can help validate whether local operational partner staff understand and are ready to leverage escalation protocols when a fleet or vehicle incident occurs that meets established thresholds.

[8] SAE J3016 defines this as "Operating conditions under which a given driving automation system, or feature thereof, is specifically designed to function, including, but not limited to, environmental, geographical, and time-of-day restrictions, and/or the requisite presence or absence of certain traffic or roadway characteristics." [15]

Additionally, exercises with government entities involved in city-wide emergency response can help establish how the AV organization and government agency will maintain situational awareness and share information during a large-scale incident affecting large portions of a city (e.g., a natural disaster). These exercises can also clarify how the AV organizations should participate in a city's central command for the incident (e.g., its Emergency Operations Center).

Across all such entities, AV companies may consider incrementally increasing the scope of each exercise using the range of exercises outlined in Section 3.2.

## 3.1 An All-Hazards approach to exercises

Incidents, by their very nature, may involve unexpected variables and the application of response protocols in novel contexts. An all-hazards approach, "an integrated approach to emergency preparedness planning that focuses on capacities and capabilities that are critical to preparedness for a full spectrum of emergencies or disasters," [54] builds incident management capability to handle this novelty. This approach focuses on developing broad capacities and capabilities [54] necessary to manage a wide range of potential incidents, rather than only building dedicated processes for specific threat scenarios.

In the United States, all-hazards approaches to incident management began to emerge in the 1970s and today are enshrined in federal doctrine and training as well as the practices of many states and localities [60,61]. Organizations across the globe also use all-hazards approaches – from the EU, whose preparedness union strategy is based on an "integrated all hazards approach," to the World Health Organization, which is developing an AI-powered all-hazards toolkit [62], and the United Nations Office for Disaster Risk Reduction (UNDRR), which employ an all-hazards approach in global framework for disaster risk reduction, the 2015-2025 Sendai Framework For Action [63,64].

These offer guidance on how to use exercises to build capabilities. Exercises are important for all-hazards capabilities because they provide an artificial environment where IM practitioners can validate if response systems including procedures, technology, tools, and personnel can be applied across multiple hazard types, permutations, cascading events, etc. In reflecting an all-hazards orientation, exercises should be designed to build overarching incident management competencies in a hazard-agnostic manner. However, to prepare practitioners for hazards most relevant to their jurisdictions, they should also focus on the hazards that are reasonably foreseeable to occur in a given locality [65].

## 3.2 Types of AV exercises

There are many different kinds of exercises. For AVs (Figure 2), these can be categorized according to what each is typically designed to achieve (although this is flexible).[9] As with other choices when planning the exercise, the type of exercise must be aligned with the objective (discussed in section 4.1.2).

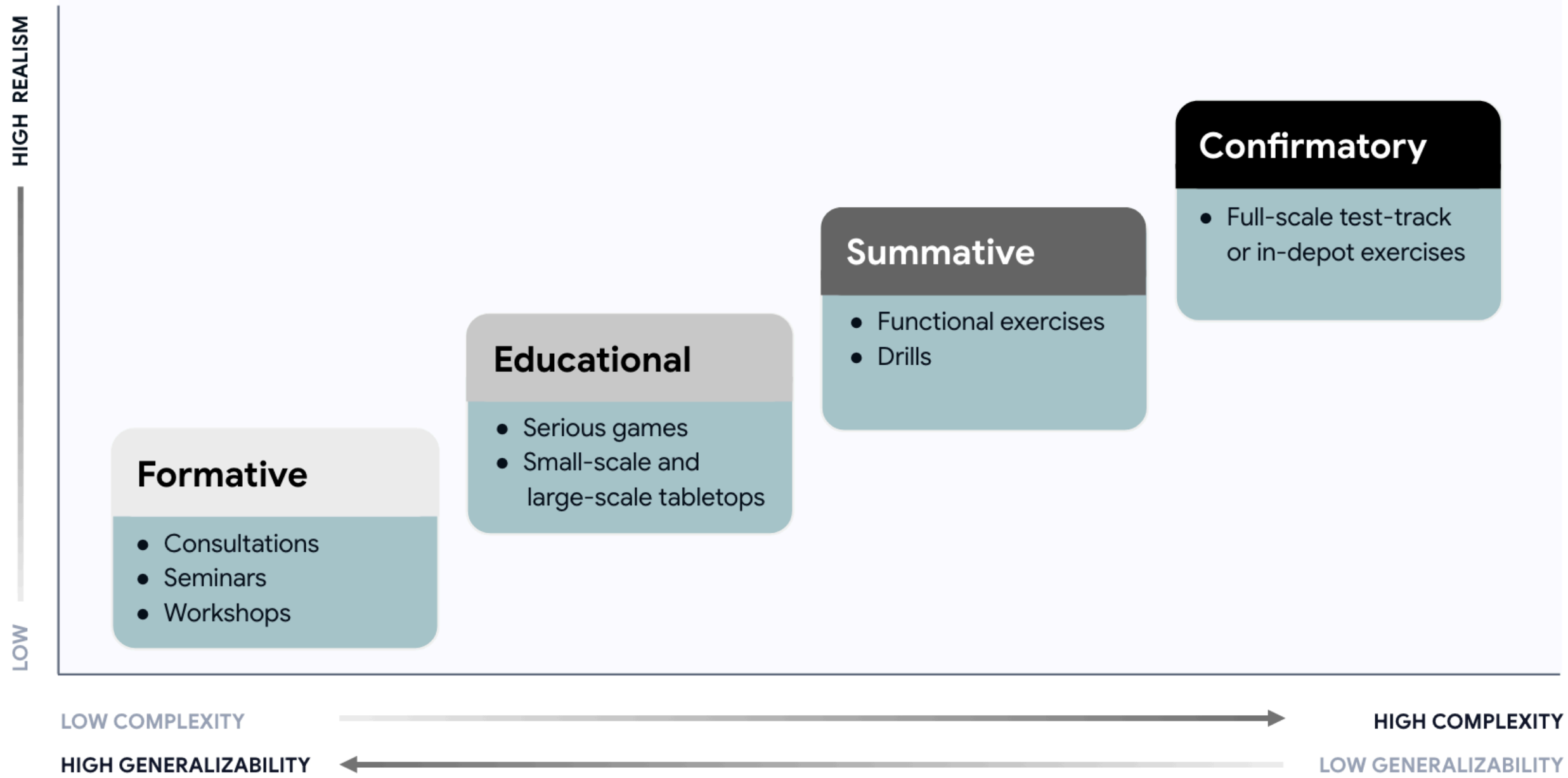

Figure 2. Types of exercises (as adapted for an AV-specific application from HSEEP's progressive exercise continuum as described by the Transportation Research Board (TRB) [55])

**Formative exercises** aim to gather input from participants and contributors and increase detailing for IM policies and procedures on specific risks. These exercises focus on plan creation,revision, and socialization [55].

**Educational exercises** focus on refining plans and teaching them to a wider audience. This may include expanding from those who set policies and procedures (i.e., participants in formative exercises) to those who contribute to incident response [55].

Educational exercises include tabletop exercises, wherein "participants are encouraged to discuss issues in depth and to develop decisions through slow-paced problem solving rather than rapid, spontaneous decision making that occurs under actual or simulated emergency conditions" [55]. These "offer the opportunity to strengthen crisis management capabilities as relatively low-cost methods of honing existing skills and identifying improvements in frameworks, plans, and policies" [66].

---

[9] For example, a tabletop is frequently educational, but can also have formative, evaluative, or confirmatory elements depending on how finely described and nuanced the scenario is, what is expected of participants, and what is done with exercise results.

Serious games, also known as war games [67], usually involve more detailed scenarios and a great emphasis on decision-making under uncertainty. They may also use game-mechanics such as random selection of complications, assignment (or removal) of resources, or even adversarial environments [67]. Role-play may also be used to spread understanding of teammates' responsibilities [68].

**Summative exercises** center on performance assessment [69]. What exactly is being assessed varies based on need. The elasticity of drills and functional exercises means that not just practitioner knowledge, but also decision-making skills, communication, information sharing, resource management, coordination, sufficiency and resilience of supporting technological infrastructure, the availability of appropriate physical resources and allocation of assets, etc. [69]. The chief distinction between drills and functional exercises is the former focuses on one specific team, aspect or process [66], while the latter covers a range of teams and processes, allowing for an assessment of not just the organizations involved but also identification of any inconsistencies or misalignments in the seams [55].

**Confirmatory exercises** substantiate views on readiness developed during previous exercises.[10] The difference between Confirmatory and Summative is not found in results; one can confirm readiness or discover new insufficiency during either. The difference instead is in the maturity of the IM program. Because full-scale exercises require the most resources [55,66], they are generally not undertaken until the IM program and the associated AV service are sufficiently advanced and have developed an exercise objective that cannot be satisfied any other way (*e.g.*, validating end-to-end response readiness for a given hazard type).

Overall, **formative** and **educational exercises** are primarily focused on *discussing* and *identifying* key response teams and tasks, while **summative** and **confirmatory exercises** share an emphasis on *executing* and *evaluating* the effectiveness of these teams and tasks under realistic conditions.

As Figure 2 shows, complexity increases as exercises shift from formative to confirmatory. So too does event realism: the extent to which exercise activities replicate closely real world conditions. However, the all-hazard application of the activity (the ability to relate exercise results to a broader range of circumstances beyond those tested) may decrease with increasing complexity and realism (both of which thrive on specificity). Additionally, because formative and educational exercises may use a wide variety of personnel but summative and confirmatory exercises often focus on a specific team or teams, results from the latter may not be generalizable to non-participating teams.

It is not possible to enact every type of exercise for every conceivable scenario (ergo the importance of the all-hazard approach); there is always a new scenario or twist on a previously enacted scenario. However, skipping an exercise category entirely may risk failing to identify major vulnerabilities and opportunities to build programmatic strength.

[10]Confirmatory exercises could potentially occur on-road, in close coordination with first responders and Departments of Emergency Management. However, this may engender considerable risks, only some of which are manageable, and the gathered intel may also be available from other, less risky settings.

# 4. Stages in an exercise

We distinguish between five stages when preparing, facilitating, and evaluating an exercise. These stages are adapted from standardized exercise frameworks. Specifically, the below content draws heavily from guidance in HSEEP, and from Van Haperen, 2001 [46]; Angafor et al., 2023 [47]; Mäses et al., 2021 [10]; Crichton & Kelly, 2012 [49]; and Van Niekerk et al., 2015 [50]

| | Situational awareness & information sharing | Incident action & implementation planning | Resource management & mobilization | Coordination & collaboration | Feedback & continuous quality improvement |
|---|---|---|---|---|---|
| **STAGE 0: Preparing the exercise** | • Create the initial information & injections participants will receive<br>• Based on objectives, identify awareness factors & sharing pathways to be assessed | • Determine needs & objectives<br>• Determine exercise type<br>• Determine scenario & the associated exercise plan | • Based on the exercise type, scenario, & objectives, determine & secure the availability of needed resources<br>• Based on objectives, identify resource management to be assessed | • Based on the exercise type, scenario, & objectives, enact any needed pre-exercise coordination | • Prepare for post-exercise evaluation by identifying the quantitative & qualitative metrics that will be used to assess whether the exercise accomplished its objectives |
| **STAGE 1: Standing up the team** | | | • Mobilize the initial team intended to respond to the event | • If the notification of an event is not automatic, enact notification protocols of appropriate personnel based on roles & responsibilities | |
| **STAGE 2: Assessing & responding** | • Determine information availability & fill gaps where necessary<br>• Maintain situational awareness<br>• Share information where appropriate<br>• Maintain documentation | • Identify & evaluate possible courses of action initially & over time<br>• Optimize based on role & priorities<br>• Clearly communicate decisions to appropriate personnel<br>• Modify plans as needed | • Mobilize & adjust staff, resources, & intelligence as needed<br>• Track all outstanding requests<br>• Maintain participants' safety<br>• Use handbooks & official guidance documents where appropriate | • Establish & maintain roles & responsibilities<br>• Maintain awareness of & act vis-a-vis external entities<br>• Escalate where appropriate | |
| **STAGE 3: Closing the exercise** | | • Decide that the exercise is complete (where appropriate)<br>• Denote next steps participants may undertake | | | |
| **STAGE 4: Evaluation** | • Evaluation of situational awareness & information sharing | • Evaluation of incident action & implementation planning | • Evaluation of resource management & mobilization | • Evaluation of coordination & collaboration | • Identify & follow up on areas for improvement in organization-wide processes, procedures, &/or performance<br>• Identify areas of improvement for future exercises |

Fig 3. Stages in the exercise process (source: authors)

Exercises start with thoughtful detailed planning to develop the exercise (stage 0, preparing) (Figure 3). Participants are notified of an event and given instructions (stage 1, standing up the exercise), who in turn engage in the exercise activities (stage 2). This will happen for some time until the exercise is closed (stage 3) and evaluated to develop lessons learned (stage 4). For the exercise to be effective, each stage involves relying on and building the functional domains necessary for incident management [45] in different ways – using and strengthening situational awareness and information sharing, incident action and implementation planning, resource management and mobilization, coordination and collaboration, and feedback and continuous quality improvement. Together, these domains capture the entirety of IM processes and procedures over the lifespan of an exercise or event [45].

Effective exercises prescribe objectives that are appropriately tailored to address organizational response needs [7]. Applying this guidance to AVs, an effective exercise increases an organization's response capabilities in a measurable way relative to the exercise's type and built-in constraints. Utilizing a mix of exercise types (as described in Section 3) can provide a more comprehensive signal regarding an organization's readiness.

## 4.1 STAGE 0: Preparation

Good exercises spring from thoughtful, detailed planning. Although no set timeline can exist for how far in advance planning should start, sufficient time should be allocated to set precise objectives and craft an activity to meet those objectives. Note that the content listed in Stage 0 is not comprehensive but rather outlines broad considerations and activities. As an initial heuristic, AV companies may consider planning on the order of weeks to prepare for most formative or educational exercise types (as identified in Section 3.2), whereas summative or confirmatory exercises may require months. In general, an exercise requires an amount of preparation that is commensurate with the level of intended realism as well as the extent to which the exercise is testing the intersection of an AV company's technology, processes, and personnel.

For example, a few weeks or less may be needed to prepare a tabletop exercise that validates whether one or a small number of teams understand their role in detecting, escalating, and responding to a collision involving one AV. In contrast, months may be needed to prepare a full-scale collision response exercise involving a vehicle and on-site personnel at a closed-course test track, the use of internal tooling and alerting systems, multiple participants across teams, and injections that develop the scenario over time.

### 4.1.1 STAGE 0.1: Conducting needs assessment

Planning begins with a needs assessment; what outstanding need may be fulfilled by an exercise? As described in Section 3, there are a range of possible triggers to initiate the planning of an exercise. An effective needs assessment should provide insight into the potential type, size, and complexity of the exercise required although these details will not be finalized until Stages 0.5 and 0.6.

There can be different types of needs, specifically:

- Exercises at the *strategic* level speak to the decision-making processes and resource allocation among senior leadership; [66]
- Exercises at the *operational* level involve practical and applied execution of tasks and implementation of protocols; [66]
- Exercises at the *tactical* level focus on “implementing strategic decisions and managing resources”. [66]

Exercises can also span levels, as explored in the table below. For example, an exercise focused on proper escalation spans all three, while one focused on coordination between a roadside assistance group and a remote internal response team spans tactical and operational.

| Example scenarios | Operational | Tactical | Strategic |
|---|---|---|---|
| Roadside assistance conducts drills on parking and safely approaching an AV that has been involved in a collision. | Yes | No | No |
| Roadside and remote assistance teams practice coordinating with internal, remote response teams regarding a collision. | Yes | Yes | No |
| Response teams engage leadership in determining a broader set of response actions including any restrictions to operations and communications to internal and external stakeholders. | Yes | Yes | Yes |

At this stage, institutional buy-in should be secured both in terms of breadth (commitment from leadership of all teams likely to be involved in the exercise, as well as senior leadership overall) and depth (commitment from appropriate parties to provide the resources and personnel required from each team, to be determined in Stages 0.2-0.7).

#### 4.1.2 STAGE 0.2: Establish the objective

Once a need is identified, it translates into a “clear and achievable” [31] objective: what is this exercise trying to accomplish? As described in Section 2.1 for exercises, the objective should be narrowly defined and action-driven. It flows directly into deciding the appropriate exercise type, event, and organizational factors.

Depending on the objective, the exercise may also benefit from the formulation of specific performance capabilities in relation to specific tasks. These are respectively referred to as capability targets and critical tasks within HSEEP [7]. The flow from exercise objectives to capability targets to critical tasks serves as the foundation of the overall exercise plan. This document details what participants are responding to initially and throughout the exercise. It includes how the event will begin (e.g., were participants provided with advanced notice prior to the exercise, how participants will be notified and what initial information they will be given) together with an overall mapping of how participants are expected to respond to the scenario (e.g., what procedures should be activated, what escalations should occur, what decisions should be made, etc.), with flexibility for divergence. It may include a timeline for the injection of new information on the original event, information on new events, or interactions with outside entities. Each injection should be associated with a specific goal, be it to test a competency, move the exercise forward, increase realistic stress, etc. Last, the exercise plan ends with an evaluation plan created before the exercise starts, to support validity of conclusions. The plan includes when and how the evaluation will take place, what it seeks to achieve, who will be involved, and qualitative and quantitative metrics by which results will be analyzed.

#### 4.1.3 STAGE 0.3: Identify exercise roles and responsibilities

Roles and responsibilities should be clearly and cleanly defined across the planning, execution, and evaluation of an exercise. There may be wide variation according to exercise's type, scale, complexity, etc., but generally roles can be categorized as follows (largely aligned with HSEEP):

- Planning team: A sole exercise director or team of individuals responsible for identifying the objective, deciding on the appropriate exercise type, designing the scenario, and conducting the work described below in Stage 0 (section 4.1).
- Controller: responsible for overseeing the exercise (described in Stages 1-3, in section 4.2-4.4 below). The precise nature of this role may vary widely by exercise type. For example, in a consultation, the controller may be deeply involved in elicitation and discussion, acting almost as an interviewer, but for a full-scale functional exercise, the controller may just trigger the initial set-up and then only observe.
- Evaluator: responsible for observing the exercise and conducting the evaluation (described in Stage 4, in section 4.5 below). The evaluator is most effective as a neutral party, not directly in the reporting chain of participants, the planning team, and the controller so as to provide an unbiased assessment especially during larger exercises.

There may be circumstances in which multiple roles are fulfilled by the same individual or team (e.g., given constrained resources and available personnel, in smaller exercises, etc.).

#### 4.1.4 STAGE 0.4: Shape exercise scenario

An exercise's scenario (or scenarios) should be suited to the objective. It should also be plausible [66,31]. Scenarios may relate to known risks or focus on what is known as Wicked Problems: problems that are "uncertain, complex, and having no obvious solution." [70] Considerations when determining the characteristics of the incident under simulation include but are not limited to:

- **The hazard being simulated**: As noted in Section 3, AVs may be affected by many types of hazards. The nature of the event will dictate other factors such as:
    - The specific safety risk (e.g., a collision involving a AV operating post-deployment, with no company employee inside, or on-site employees who may be at risk from an active shooter);
    - The amount of warning available in advance of an event (e.g., hurricane warnings may start days in advance, tsunami warnings may be hours, earthquakes may be instantaneous);
    - The time-scale of the event (e.g., a collision occurs over the course of seconds, but severe weather can continue for an indeterminate amount of time);

- The geographic breadth of the event (e.g., a collision at one street corner versus an entire region experiencing an earthquake);
    - The complexity of the event in terms of involved parties and vehicles (e.g., one AV incident response team managing a two-vehicle collision, as compared to being one entity in a wide range of organizations, all being directed to respond to a natural disaster in key ways by a centralized government organization);
    - The location of the event (e.g., the affected AV or AVs are on a freeway or surface street).

- **The impact of the hazard being simulated:** How does the event impact the involved individuals, vehicles, organizations, and geographic locations? What is the risk in terms of severity and number of parties involved? What kind(s) of impacts could this event have, beyond safety (e.g., is there the potential for social or economic impact)? Is there risk generally or are specific types of individuals at elevated risk?

- **The AV company's direct or indirect responsibility** in managing the incident: Does the event directly fall under the AV company's area of focus (e.g., a collision), or under the purview of a contractor or collaborator (e.g., a flooded facility wherein the site is operated by a contractor)? Is the event in an area in which the AVs operate and is specific to transportation (e.g., a major set of thoroughfares being blocked due to construction) or are the effects more widespread (e.g., a tsunami or volcanic eruption, wherein an AV company may support a coordinated response as planned for by government disaster response agencies)? This defines the operator's responsibilities and abilities in terms of how much control they can and should exercise over the response.[11]

#### 4.1.5 STAGE 0.5: Select appropriate exercise type

The choice of exercise type focuses on what would satisfy objectives and align with the level of maturity of the process under examination. Succinctly, "target capabilities determine the exercise, not the other way around." [8] Some objectives (e.g., developing and testing response procedures for a hazard type faced by AVs in a new city) may require the completion of several exercises with increasing realism and complexity over time to provide incremental validation. One type of exercise is not inherently better, more useful, or more powerful than another; each serves a specific purpose.

---

[11] While many areas of stage 0 require the guidance given in programs like HSEEP to be adjusted when applied to AVs, scope of responsibility during incident response may require the closest scrutiny. Official guidance in programs like HSEEP is generally for government IM practitioners. The responsibilities, legal powers and jurisdictions, response protocols, resources, etc. may differ widely from a non-government enterprise that deploys AVs.

### 4.1.6 STAGE 0.6: Identify exercise parameters and needed resources

Several intertwined organizational considerations factor into the construction of exercises. The choice of exercise type and scenario may dictate or influence[12] these considerations, which include but are not limited to:

- **When and How Long**: How long is required to accomplish the stated objectives? Such is not an exact science, but one may not want to schedule 5 hours for a lightweight tabletop, nor a full-scale test track exercise for 15 minutes. Depending on the exercise type, the length reflects the amount of time required to thoroughly discuss and/or perform the full set of tasks involved in responding to the given scenario.

- **Where:** The medium through which the exercise is taking place. For full-scale exercises, physical space, such as a closed-course test track, must be secured. Other exercise types may warrant a hybrid format with some response activities occurring on-site while others happen simultaneously over video call or other communication mediums (e.g., real-time coordination between dispersed and on-scene personnel). Or it can be conducted virtually. In addition to providing logistical advantages, the virtual format may also mirror real-time response practices for AV companies with increasingly large geographic footprints and centralized incident response functions [71].

  Regardless of the format, the exercise should be known as a safe space: safe in terms of no physical harm coming to participants but also safe in terms of participants being able to make mistakes, ask questions, and express uncertainties without negative consequences [66].

- **With what:** Artifacts containing simulated content may be needed to lay out the initial parameters, enliven the exercise, increase engagement, and make it easier for participants to respond and make decisions. Such can increase the participants' understanding of what is happening at the scene of the event as well as how the event is affecting the broader organization and communities in which it operates – emails, video, internal or external communications including social media posts, news articles, etc. The level of provided information is dictated by objectives; if the objective focuses on decision-making under circumstances of limited intel and deep uncertainty, little information may be provided.

- **Who:** What type and extent of teams are involved? For example, is this a train-the-trainer activity, an activity with one specific response team, or an activity that will be repeated for all response teams? Similarly, as dictated by the exercise's objectives and the event's severity, does this involve one response team or is it an all-hands-on-deck situation? Additionally, an exercise may focus on a specific area or time period of response. If appropriate, other employees or consultants can be brought into role-play as outside entities. One may want to notify uninvolved employees that an exercise is occurring.

---

[12] Ideally, it is the exercise type dictating or influencing these logistical considerations, and not the other way around.

- **How many:** Is there more than one exercise occurring at the same time? If there are multiple events, is each unique or cascading events of varying natures (i.e., a Mrs. O'Leary's Cow situation)? For example an earthquake may result in vehicle strandings. This may function as a stress-test to identify single points of failure and assess the robustness of the IM processes, team dependencies and fallbacks.

Once these are all decided, two more roles can be filled:
- Participants: the individual, team, or teams participating in the exercise.
- Observers: individuals present for learning purposes but are not participating in, running, or evaluating the exercise.

#### 4.1.7 STAGE 0.7: Finalize exercise test plans and documentation required for each participant role

The final stage of preparation involves finalizing all documentation (which has been created over the course of Stages 0.1 to 0.6). This includes but is not limited to the creation of detailed exercise objectives, capability targets, critical tasks, descriptions, timetables, profiles for participants playing roles other than their own, exercise artifacts, and tools to be used in the evaluation. All documentation and artifacts should be clearly labeled as specific to the exercise.

### 4.2 STAGE 1: Standing up the team

This stage marks the beginning of the exercise. All participants are notified of an event and are given instructions of how and where to assemble.

Pursuant to exercise objectives and type, this stage may test notification software and protocols within the company, as well as response behavior of participants.

### 4.3 STAGE 2: Assessing and responding

During this stage, participants assess the situation and respond accordingly. The exact nature of this stage may vary widely pursuant to the scenario.

Activities within specific IM domains may include [45] but are not limited to those described in Figure 3.

### 4.4 STAGE 3: Closing the exercise

The decision to end an exercise is complex. First, there is the consideration of on what basis an exercise ends. It could be that all possible actions have been taken, or next steps require information that is not currently available or require days-long individual work. It could be that the exercise simply runs to its allotted time or is determined to have achieved its objectives. Second is the question of who decides the exercise is over (i.e., the controller, the participants, or another entity).

Arguably, any of these options are reasonable, pursuant to the objectives, exercise type, and scenario. What is essential is that the parameters of the decision should be identified in Stage 0 rather than during the exercise.

### 4.5 STAGE 4: Evaluation and Continuous Improvement

Evaluation has twin purposes: to analyze the actions undertaken by participants to improve processes, procedures, and/or performance, and to analyze the quality of the exercise itself to enhance future exercises. Evaluation aims to understand if the right activities resulted in the right outcomes at the right time, with the right speed, and in the right order.

Evaluation may start at any stage in the exercise.[13] Immediately following the exercise a hotwash (defined as a "meeting that provides an opportunity to discuss exercise strengths and areas for improvement immediately following the conduct of an exercise") with participants may occur and can be facilitated by the controller, the evaluator or an informed observer [7]. Controllers, evaluators, and the planning team should also hold a debrief to review their observations. Time for debriefing and hotwashing is part of scheduling the exercise; allotting insufficient time for these important activities is a common mistake [66].

Evaluation focuses on results specific to each exercise objective. HSEEP's framework for evaluation, which hinges on clearly identifying each exercise objective and its corresponding set of capabilities and capability targets, is useful here. By measuring whether the tasks involved in each capability target are fulfilled in the desired timeframe, evaluators using the HSEEP framework can assign performance ratings per objective that indicate the extent to which the objective was achieved and the significance of any identified opportunities to improve overall response [7]. This approach is particularly beneficial when evaluating a summative or confirmatory exercise that is designed to validate a specific set of response actions and outcomes.

Beyond assessing performance for specific objectives, exercises are opportunities for discovery. Evaluation should presuppose as little as possible and evaluators should be open-minded to the observation and identification of previously unidentified or undetected weaknesses, assumptions, dependencies, uncertainties, etc. To support this, evaluators may consider observations across IM domains as found in Nelson et al [72] to identify lessons learned.

An exercise's scope, type, and objectives determine the evaluation's qualitative and quantitative data sources. Common sources involve gathering feedback from exercise participants, observers, and evaluators through:

- Semi-structured interviews with individuals and groups who participated, and with key observers;
- Structured debriefs and hotwashes;
- Review of exercise documentation and evaluation guides;

---

[13] Pausing the exercise periodically to allow for reflection is also an option [66], but can diminish realism and prevent mistakes from playing out, an important part of the simulated activity.

- Structured or open-ended questions within anonymous surveys.

The evaluator's own observations may provide additional supportive or conflicting intel. For example, the evaluator may track desired versus realized response time or the percentage of key tasks completed as part of an objective's capability. Additional data from beyond the exercise may also be used. For example, if a similar, real event occurred, results may be compared. Alternatively, if the exercise is one in a series of activities over time, cross-exercise results may be assessed to identify trends.

Evaluation depth and robustness may also vary across exercise types. For example, the evaluation of a summative or confirmatory exercise will continue into subsequent weeks and draw upon a wider range of data sources. Evaluators may wish to choose multiple tools to enhance validity.

The exercise type also influences who does the evaluation. Ideally an evaluator balances independence and expertise. This can be challenging because, notably in the earlier stages of an IM program development, the expertise in IM, ability to run an exercise, understanding of how to evaluate, and standing to act on results may all reside within the same small group or even the same individual. Section 4.1.3 discusses requirements for the evaluator role, including neutrality, particularly important for Summative and especially Confirmatory exercises. Depending on the objectives, additional evaluators with narrow subject matter expertise may also be included.

Conclusions from the evaluation are discussed in an After-Action Meeting (AAM) and are documented in the AAR. These forums provide an answer to the questions of "what happened? What was supposed to happen, based on current plans, policies, and procedures? Was there a difference? What was the impact?" [7] Content may include, but is not limited to:

- Summarizing the exercise;
- Reviewing results of the evaluation, both in terms of actions undertaken (or not undertaken) and exercise quality and performance rating per objective;
  - Considering how these results align with real events and/or other exercises;
- Suggesting areas of improvement, and specific action items to be undertaken to support continuous quality improvement in real-world responses;[14]
  - Advising on prioritization of actions;
  - Articulating the likely consequence of not undertaking these actions;
  - Considering how the proposed action aligns with or diverges from long-term pre-existing plans.

With some minor changes to roles, the evaluator may choose to use the HSEEP template when generating the AAR [73], an alternative source [74], or create their own.

---

[14] As a stark example of what can happen when the conclusions of exercises are not heeded, a year before 2005's Hurricane Katrina, an exercise for "Hurricane Pam" focused on a hurricane hitting New Orleans. However, "admonitory lessons were either ignored or inadequately applied" [75]. This contributed to the inadequate preparedness and limited response [75] for Hurricane Katrina. Similar circumstances occurred with the Cygnus exercise which, in 2016, presaged issues that arose in the 2020 pandemic [66].

Below is a high level example tabletop exercise scenario involving an AV operator and public safety partners in San Francisco followed by a hypothetical approach to evaluation.

### *4.5.1 Example scenario:*

At 14:00 (2:00 PM) on a Tuesday afternoon, a magnitude 7.0 earthquake strikes the Bay Area, originating from the Hayward Fault with an epicenter located just east of Oakland. The intense ground shaking lasts for approximately 45 seconds, causing widespread and varied damage across the City and County of San Francisco and the East Bay.

- **Infrastructure Impacts:** There is significant, yet localized, damage to transportation infrastructure. Initial reports indicate compromised overpasses, localized structural failure on key city arterial roads, and widespread debris and possible liquefaction near coastal areas. The full extent of road network damage is currently unknown.
- **Communications Disruption:** Cellular service is severely degraded or intermittent across large parts of San Francisco and the East Bay due to power outages and tower damage, challenging the AV operator's ability to maintain reliable, real-time connectivity with its fleet.

**Objective:**
Assess the ability to establish and maintain timely direct communications in support of situational awareness and ongoing operations among and between public safety partners and an AV operator's incident response team.

**Evaluation:**
Prior to the exercise, the evaluator already established the set of capabilities and corresponding tasks and processes that would need to be executed in order to accomplish the objective.

- During **Stage 1,** the evaluator measures that, among other considerations:
  - The AV operator team establishes contact with city first responders within X minutes of the earthquake occurring and joins the city's Emergency Operations Center (EOC) within Y minutes of the EOC convening.
- During **Stage 2,** the evaluator determines that, among other considerations:
  - While both involved teams successfully cited and demonstrated understanding of 3/3 relevant processes, a few participants expressed some uncertainty regarding the desired service level agreement (SLA) for completing 1 process.
- During **Stage 3:** the evaluator determines, among other considerations, that the exercise is ended at the appropriate time or after the participants have completed all response tasks.
- During **Stage 4,** the evaluator also considers inputs from the post-exercise hotwash and anonymous survey for observers and participants. During an additional debrief and interview, a participant raised concerns around timing for a specific process. In the survey, 20% of respondents provided a rating of 4 when answering a 1-5 Likert scale question about clarity of each participating team's role in responding to the scenario with the other 80% of respondents responding with a rating of 5.

The evaluator synthesizes these inputs and, in corroboration with their own findings and assigns a performance rating of Performed with Some Challenges (S) according to HSEEP's ratings definitions [73] which indicates that participating teams achieved the exercise's intended objective but that an opportunity to enhance effectiveness and/or efficiency was identified. Had the participants not demonstrated understanding of all 3 processes or if survey or hotwash feedback indicated other significant gaps in process or understanding, the evaluator would have adjusted their rating to either a Performed with Major Challenges (M) or Unable to be Performed (U) according to the HSEEP rating scale [73].

The evaluator outlines these findings orally in the AAM and in writing in the AAR including the process that needs to be improved and a specific action item assigned to the appropriate owner to clarify the desired SLA and then re-train the involved team.

# 5. Exercise opportunities and limitations

AV exercises face the same limitations as other IM exercises. These inherent limitations can lead to a false sense of security if not properly managed. There is the old adage in IM: if you have seen one incident, you have only seen one incident. Even using an all-hazards approach, limits exist to what can be transferred from any exercise to similar events in the real world.

The primary limitations of exercises fall into two categories: structural artificiality and psychological factors. Regarding the first, even the most realistic exercises face limitations on realism. Such may reduce an exercise's ability to demonstrate the difficult trade-offs often required in real-world AV incident response. Additionally, this artificiality constrains decision making and improvisation [76]. In response, exercises may use historical data and hypothetical incident scenarios as bases. Ownership of historical operational and vehicle data may present opportunities for AV operators to create engaging and realistic exercise scenarios.

Regarding the second limitation, the stresses of an exercise's artificial environment may not mirror the intensity of real-world response and can mask the evaluation of performance under pressure. Participants may also fight the scenario, identifying flaws in the exercise design versus reacting to injections (i.e., new information) as presented. Hiring staff familiar with IM protocols and exercise design can help bridge this gap. Additionally, as noted in section 4.1.4 and 4.1.6, detailed planning, coupled with realistic artifacts, is particularly important. Additionally, respect, professionalism, and engagement for these activities by participants, role-players, and observers are a key element of a company's safety culture.

Despite these limitations, IM exercises remain one of the most potent tools for increasing an AV company's overall response capabilities. AV IM exercise practitioners can minimize the effects of any one exercise's limitations by:

1. Practitioners may strive to increase the quality of each exercise using the recommendations outlined in sections 4.1-4.5;

2. Practitioners may increase exercise diversity and volume. Exploring the same or similar scenarios through different exercise types over time (formative, educational, summative, confirmatory) can help to provide a more balanced picture of the organization's readiness.

Even the most effective exercise systems are only one part of an organizational approach to AV safety. Other significant benefits may be derived from further investment in areas such as the organization's safety culture and frameworks (e.g., Safety Management System), and mechanisms for learning and improving following real incidents including creating AAR and identifying and resolving corrective actions.

# 6. Discussion and Conclusions

Exercises are a critical tool for building robust AV IM and response processes. Thoughtful and detailed exercises support AV companies in being prepared to efficiently and effectively respond to a vast spectrum of hazards.

The discussed frameworks balance specificity and generalizability by aligning with general IM practices while focusing on the unique elements of AVs. To support IM through exercises, an AV company should consider:

- [section 2] Fostering a workforce with both deep and broad subject matter expertise across a range of hazard types and IM disciplines;
- [section 2.1] Positioning exercises as a key component of an effective AV IM program;
- [section 3] Ensuring institutional commitment to the IM program generally and to exercises specifically; an AV IM exercise program should grow in coordination with the company's maturity;
- [section 3.1] Using exercises to validate and improve an all-hazards approach to IM, motivated by the known impossibility of covering all possible hazards (and permutations and combinations), regardless of resources;
- [section 3.2] Cultivating capabilities and resources to utilize different exercise types and determine which exercise type aligns with a given objective;
- [section 4, 4.1, 4.1.1 - 4.1.7] Maturing internal abilities in exercise development;
- [section 4.2-4.4] Supporting, through time and resources, the stand-up, enactment, and closing of an exercise, the activities of which span all IM domains;
- [section 4.5] Analyzing the results of an exercise to identify strengths and gaps, both in participant actions and in the exercise itself;
- [section 5] Minimizing exercise limitations through increasing exercise quality, diversity, and volume; building capacity in skills and topics which exercises cannot support.

## 6.1 Future work

As organizations adapt and use this paper's content, efforts should focus on refining, validating, and improving its components. Exercises are a living discipline, for which this paper has established a baseline for AV IM practitioners. We anticipate further development as the sophistication of AV IM deepens, into exercises specifically and into IM broadly. Additionally, there may be opportunities in standards bodies to create general practices around, for example, exercise design or AARs specific to AV IM exercises.

Last, this paper was limited by scope, focusing on a fleet-based model for L4 and L5 AVs. Although there are clear implications for L3 vehicles and for privately owned business models, these may each require different procedures and may give rise to different responsibilities.

Similarly, while some of this material may be broadly applicable, much of the literature is U.S.-focused. Additional research is needed to identify and incorporate potentially unique international and multi-country considerations for AV IM exercise practitioners.

## 6.2 Conclusions

This paper tailored existing research and IM exercise practices for AVs, an activity necessitated by the availability of general guidance, lack of specific guidance, and unique socio-technical environment which has grown around and is shaped by AVs as an emerging technology within private industry [77].[15] Frameworks in this paper embed common IM processes and protocols, improving intercompany and inter-organization coordination and collaboration by supporting a common understanding, and by enhancing anticipation and legitimacy [78,35].

For any transportation provider, responding to events is essential and inevitable. The value proposition of exercises is to practice readiness in prevention, response, and mitigation. Developing an effective AV exercise program using guidance from this paper can de-risk AV operations and build an organization's capability to efficiently and effectively address events when they arise.

[15] Research on emerging technologies like AVs suggests that risk management strategies should be tailored to the associated socio-technical environment. Similar views can be found in risk management for electric grid cybersecurity risks [79,80] and the implementation of smart city technologies [81].